\documentstyle[]{mn}
 
\title[High Resolution HI Study of NGC4151]{Gas Dynamics in the Barred Seyfert Galaxy
NGC4151 - II. High Resolution HI Study}
 
\author[C. Mundell et al.]
        {C.G. Mundell$^{1}$\footnote{Present address: Department of Astronomy, University of Maryland, College Park, MD~20742, USA.}, A. Pedlar$^{1}$, D.L. Shone$^{1}$, A. Robinson$^{2}$
                \\
$^{1}$University of Manchester, Nuffield Radio Astronomy
Laboratories,
Jodrell~Bank, Macclesfield, Cheshire SK11~9DL, UK.\\
$^{2}$Division of Physical Sciences, University of
Hertfordshire, College Lane, Hatfield, Hertfordshire, AL10 9AB, UK.\\
$^{\star}$Present address: Department of Astronomy, University of Maryland, College Park, MD~20742, USA.}

\date{}
\date{Accepted for publication in MNRAS}
\begin{document}
\maketitle
 
\begin{abstract} 
\large

{We present sensitive, high angular resolution ($6''\times5''$)
$\lambda$21-cm observations of the neutral hydrogen in the nearby
barred Seyfert galaxy, NGC4151. These HI observations, obtained using
the VLA in B-configuration, are the highest resolution to date of
this galaxy, and reveal hitherto unprecedented detail in the
distribution and kinematics of the HI on sub-kiloparsec scales.  A
complete analysis and discussion of the HI data are presented and the
global properties of the galaxy are related to the bar dynamics
presented in Paper I.

HI absorption, consistent with previous studies, is detected against
the radio continuum nucleus and shows two components -- a deep
absorption component, centred at 987 $\pm$ 1 km s$^{-1}$ and width 87
$\pm$ 3 km s$^{-1}$, and a weaker component, redshifted to 1096 $\pm$
6 km s$^{-1}$ with a width of 35 $\pm$ 15 km s$^{-1}$. An alternative
fit is also presented.  In addition to the absorption, a high velocity
cloud is detected in emission, coincident with the nucleus. This cloud
is red-shifted by 260 km s$^{-1}$ from systemic, has an HI mass of 2.3
$\times$ 10$^7$ M$_\odot$, and corresponds to outflow on the far side
of the nucleus.

Contrary to previous studies, no HI bridge is detected reaching from
the shocks directly across the nucleus. Instead, the gas streams from
the shocks onto smaller orbits and forms fingers of HI which wind {\em
around} the nucleus, consistent with predictions from general
numerical simulations of bars. These fingers correspond closely with
dust arcs seen in optical studies and resemble nuclear features seen
by others in weak barred galaxies such as M100.

A new rotation curve is presented, extending to within 8$''$ of the
nucleus and showing a turnover at a radius of $\sim$35$''$, which was
previously undetected in lower resolution studies.  The corresponding
resonance curve and the properties of the shocks (Paper I) yield a bar
pattern speed of 24.5 $\pm$ 3.7 km s$^{-1}$ and one Inner Lindblad
Resonance (ILR) at a radius of 2.8 $\pm$0.6 kpc.  Our observations,
however, do not rule out the possibility of an Inner ILR.}
\normalsize

\end{abstract}


\section{Introduction}

Neutral hydrogen is present on a wide range of scales in galaxies,
often extending far beyond the optical extent of a galaxy (Broeils \&
Rhee, 1997 and references therein), and is an important tracer of
galactic structure and dynamics.  The presence of HI in a galaxy is
related to the formation of many observed features; dissipation in the
gaseous ISM enables the formation of transient spirals, bars and other
large scale structures (Schwarz, 1981; Palou\v{s}, 1997).  Seyfert
galaxies are the closest, most common type of Active Galactic Nuclei
(AGN) and are often gas-rich, making them useful laboratories for the
study of the AGN phenomenon and its relationship to the gaseous medium
in the host galaxy.  On the largest scales, HI is sensitive to
interactions and tidal features may extend up to 200kpc (e.g. Simkin
et al., 1987; Mundell et al., 1995a). On intermediate scales spiral
features and galactic bars are well traced in HI and on sub-kpc scales
HI may play a role in the fuelling of AGN and in turn be affected by
the nuclear activity.

Nuclear activity in galaxies, either as AGN or as starbursts, requires
a ready supply of fuel and an efficient mechanism for its delivery to
the site of activity (Shlosman \& Noguchi, 1993).  In order to fuel an
{\em AGN}, the gas must lose sufficient angular momentum to reach the
very central accretion radii without triggering star formation, which
would inhibit accretion processes (Barnes \& Hernquist, 1992). To date
this has been simulated quite successfully down to scales of $\sim$1
kpc but significant momentum losses on smaller scales have been
difficult to achieve.

The rate of accretion or formation and growth of the central black
hole itself may be regulated by processes in the host galaxy, such as
galactic disk instabilities at different epochs (Shlosman \& Noguchi,
1993).  Bars, which may be tidally triggered by galaxy interactions
(Barnes \& Hernquist, 1992) or formed by gravitational disk
instabilities, are thought to be an efficient mechanism for
transporting galactic material into the central regions of galaxies
(Larson, 1994; Wada \& Habe, 1995), and for dissipating angular
momentum via shocks in the bar (Athanassoula, 1992b). Many nearby barred
spiral galaxies possess central mass concentrations with associated
non-circular motions and Gerhard (1992) suggests that radial gas
inflow is a ``common phenomenon''.  A subtle balance however, must be
maintained such that the fuel is delivered to the AGN without
accumulating so large a central mass concentration that the bar itself
is destroyed (perhaps 1--5\% of the galaxy mass concentrated in a
dense core -- Hasan \& Norman, 1990, Hasan, Pfenniger \& Norman, 1993,
Friedli, 1994, Sellwood, 1996). The accumulation of a central mass causes
the periodic orbits in the bar to become increasingly chaotic, leading
to the destruction of the bar and hence the supposed fuelling
mechanism.

Statistical studies of the correlation between bars and Seyfert
activity are controversial (Arsenault, 1989; Ho, Filippenko \&
Sargent, 1997; Heckman 1980; Simkin Su and Schwartz, 1980; Moles et al
1995). However, there is increasing evidence from near-infrared
studies that as many as 70\% of galaxies may possess bars (Mulchaey,
Regan and Kundu, 1997), although a significant number of Seyfert
galaxies do not (Mulchaey \& Regan, 1997). The situation is somewhat
complicated with the possibility that those AGN without a bar now, may
have had a bar in the past, which has been destroyed by the
accumulation of the central mass. (Such Seyferts may lack the gaseous
ISM needed to 'cool' and maintain the bar).  Detailed observations and
theoretical studies of bars in AGN complement such statistical studies
in an attempt to understand the physical processes at work.

Perhaps one of the most famous and best studied Seyfert galaxies is
NGC4151. This galaxy, and its nucleus, has been studied extensively at
all wavelengths and was one of the original Seyferts noted in the
1943 sample (Seyfert, 1943) of galaxies with bright star-like nuclei
and broad emission-line spectra. NGC4151 has a Seyfert type 1.5
nucleus (Osterbrock and Koski, 1976) in an (R')SAB(rs)ab host galaxy (de
Vaucouleurs et al., 1991). Over one thousand papers have been published on
NGC4151 and a detailed assessment of the observational status of
NGC4151 at all wavelengths has been presented in an excellent review
by Schultz (1995).

The most recent classification of NGC4151, that of type (R')SAB(rs)ab
(de Vaucouleurs et al., 1991), takes into account the well-defined two-arm
outer spiral pattern and the central bulge or `fat' bar. Initially,
however, NGC4151 was believed to be a small spiral galaxy, inclined at
37$^{\circ}$ with its major axis in PA $\sim$130$^{\circ}$. This
description arose because the optical emission originates
predominantly from the active nucleus and the central fat bar (a small
string of HII regions situated near the each end of the bar looked
like small spiral arms). Early HI observations (Davies, 1973),
however, revealed a much larger outer spiral structure that suggested
a more face-on inclination of $\sim$20$^{\circ}$ and a major axis PA
of 26$^{\circ}$. These outer spiral arms begin at each end of the bar
and wind out to a radius of $\sim$6$'$. The spiral arms are
particularly gas-rich and, along with the bar, are clearly visible in
subsequent HI images by Bosma, Ekers \& Lequeux (1977) and Pedlar et
al. (1992). The outer arms are faintly discernable in Arp's (1977)
deep optical exposure of the galaxy but it is surprising that such a
high concentration of neutral gas produces so little star formation.

NGC4151 is interesting as both an AGN and a barred galaxy and in this
paper, we address both the physical properties which relate to the bar
and outer spiral structure, and which are of general interest in the
study of barred galaxies, and the role which the bar may play in the
fuelling of the active nucleus.

We present here a complete analysis and discussion of the highest
resolution HI emission-line data of NGC4151 to date. The HI
distribution and kinematics of the host galaxy, imaged with the NRAO
Very Large Array (VLA\footnote{NRAO is operated by Associated
Universities, Inc., under cooperative agreement with the National
Science Foundation.}) in B configuration, are presented and extend the
detailed lower resolution (17$''\times$22$''$) study by Pedlar et
al. (1992). Key observational results, which have been described in
Paper I (Mundell \& Shone, 1998), are related to the global properties
of the galaxy, and the implications for fuel transport to the AGN are
discussed.

The heliocentric radial velocity of $\sim$1000 km s$^{-1}$ is often
used to estimate the distance to NGC4151. However, several authors
have noted that by including a correction for the Virgocentric flow,
in addition to the galactocentric correction, the distance estimate
can increase by almost 50\%. Tully (1988) obtained a distance to
NGC4151 of 20.3 Mpc (H$_0$=75 km s$^{-1}$ Mpc$^{-1}$) after applying
his Virgocentric correction. More recently, when Aretxaga \&
Terlevich (1994) took into account the relative velocity of the local
group with respect to the Virgo cluster they obtained a corrected
recession velocity of 1330 $\pm$ 200 km s$^{-1}$, yielding a distance
of 17.7 Mpc (H$_0$ = 75 km s$^{-1}$ Mpc$^{-1}$). So for 50 $<$ H$_0$
$<$ 100 km s$^{-1}$ Mpc$^{-1}$ and 1000 $<$ V $<$ 1523 km s$^{-1}$,
the distance lies in the range 10 $<$ D $<$ 30.5 Mpc. The uncertainty
in the value of H$_0$, however, is as large as the Virgocentric
correction (Schultz 1995) so for the purposes of this paper we
estimate the distance to NGC4151, using the heliocentric velocity of
998 km s$^{-1}$ and H$_0$=75 km s$^{-1}$ Mpc$^{-1}$, to be 13.3 Mpc. One
arcsecond corresponds to 65pc in the galaxy at this distance.

\section{Observations and Reduction}

The $\lambda$21-cm radio observations of NGC4151 were obtained in
April 1993, using the Very Large Array (VLA) in B
configuration. Spectral line mode 1A was used, with automatic
band-pass calibration and on-line Hanning smoothing performed on the
data.  Data suffering from `shadowing' (where one antenna's line of
sight is obscured by another) were discarded.

Observations of the point source 1225+368 were interleaved through the
source observations throughout the observing run to enable phase
calibration of the data. The absolute flux scale was determined from
observations of 3C286, assuming a flux density of 14.755 Jy at 1415
MHz (Baars et al. 1977).  3C286 was also used to determine the
bandpass corrections in the spectral line data.  The observing
parameters are given in Table \ref{obsparms}.

All data processing was performed using the NRAO AIPS (Astronomical
Image Processing System) package. The calibration and initial data
editing were carried out using the `pseudo-continuum' channel (channel
0) which is formed on-line by averaging the central 75\% of the band
(4.7 MHz).  The resulting phase and gain solutions were then applied
to the B array spectral line data. These data were then Fourier
transformed to produce two cubes, one using natural weighting to
optimise surface brightness sensitivity, the other uniformly weighted
for maximum resolution. The cube dimensions were 512 x 512 x 64, with
2$''$ pixels.

Continuum images were formed by averaging 10 channels, free of line
emission (channels 10-15 and 55-60 ), in the natural and uniform
cubes, avoiding channels at the ends of the band. The dirty continuum
images were subtracted from the corresponding dirty spectral line
cubes, to form continuum free cubes.

The continuum images and continuum-free cubes were then deconvolved
using the CLEAN algorithm (H\"ogbom, 1974), to remove sidelobe
structure. The images produced from the naturally and uniformly
weighted data were restored with effective beams of 6 $\times$ 5
arcsec and 4.71 $\times$ 4.71 arcsec half-power full width (FWHM)
respectively.

\setcounter{table}{0}
\begin{table}
\centering
\caption{Observing Parameters for VLA B array observations of NGC4151}
\label{obsparms}
\begin{tabular}{lc} 
\hline\\
 {\bf GALAXY}  &{\bf NGC4151} \\
 Date of Observations & April, 1993\\
 & \\ 
Field Centre  & $\alpha$= $12^h08^m01.0^s$\\ 
(B1950) & $\delta$= $39^{\circ}40'60.0''$\\
 &  \\
Central Velocity & 1000 km s$^{-1}$\\ 
No. of Antennas & 27\\ 
& \\ 
Time on Source &  8.7 hours\\
 & \\
Frequency Channels & 64 \\ 
& \\ 
Total Bandwidth &  3.125 MHz\\
  &  \\ 
Channel Separation & 48.828 kHz\\
 & \\ 
Velocity Resolution & 12.5 km s$^{-1}$\\
 & \\ 
Amplitude Calibrator & 3C286\\ 
Phase Calibrator &1225+368\\ 
&\\ 
Spectral Line Mode & 1A \\  
Hanning Smoothing  & Yes\\ 
& \\ 
Primary Beam & $30'$\\ 
& \\ 
Cellsize (arcsec) & 2\\ 
& \\
Channel No.s in squash& 10-15 \& \\ &55-60\\
&\\
\hline
\end{tabular}
\end{table}

\section{Results}
\subsection{$\lambda$21-cm Continuum Emission}

The CLEANed $\lambda$21-cm continuum image, formed from the line-free
channels (see Section 2) in the uniformly weighted cube, is shown in
Figure \ref{chanab}. The total flux density from NGC4151 is 336 mJy,
consistent with the spectral index of --0.71 found by Hamilton
(1995). The rms noise in the image is 0.26 mJy beam$^{-1}$ and the
beamsize is $4.71'' \times 4.71''$ (PA 45$^{\circ}$). The emission is
marginally resolved at this resolution and a two-dimensional Gaussian
fit to the component gives dimensions of $5.41'' \times 4.78''$ (PA
79.1$^{\circ}$). This slight elongation in PA 79$^{\circ}$ is
consistent with the presence of the $\sim$4-arcsec radio jet in PA
~77$^{\circ}$, seen at higher resolution and at other frequencies
(Pedlar et al., 1993, Mundell et al., 1995b).

\subsection{Absorption Against the Radio Continuum Nucleus}

Channel maps of the absorption feature, from the continuum subtracted
uniform cube, are shown in Figure \ref{chanab} along with the uniform
continuum image in the top left corner.  The absorption appears to be
spatially unresolved at the resolution of 4.71$'' \times$ 4.71$''$, as
the spectra do not show absorption coincident with the eastern
extended continuum emission.

\setcounter{figure}{0}
\begin{figure}
\setlength{\unitlength}{1mm}
\begin{picture}(10,90)
 
\put(0,0){\includegraphics{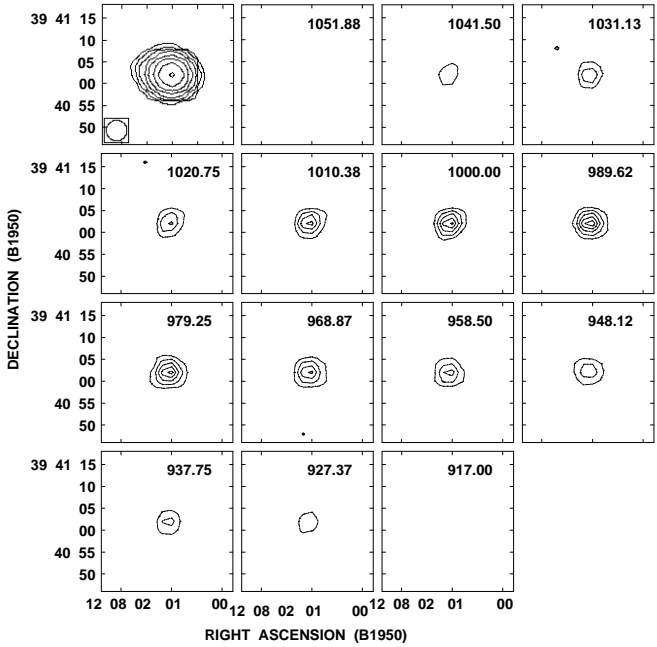}}
\end{picture} 
\caption[B array absorption maps of NGC4151]
{Maps of the channels containing absorption against the radio
continuum nucleus, from the uniformly weighted cube. The
$\lambda$21-cm continuum image can be seen in the first (top left)
panel and the beamsize for all panels is 4.71$''\times$ 4.71$''$
(shown in the lower left corner of the first panel). The continuum
contour levels are (-0.6, 0.6, 1.2, 2.4, 4.8, 9.6, 19.2, 38.4, 76.8)
mJy beam$^{-1}$. The absorption contour levels are (-10, -8,
-6, -4, -2, 2) mJy beam$^{-1}$ and the velocity (in km s$^{-1}$) of each
channel map is marked in the top right corner of each map.}
\label{chanab}
\end{figure}

\begin{figure}
\setlength{\unitlength}{1mm}
\begin{picture}(10,70)
 
\put(0,0){\includegraphics{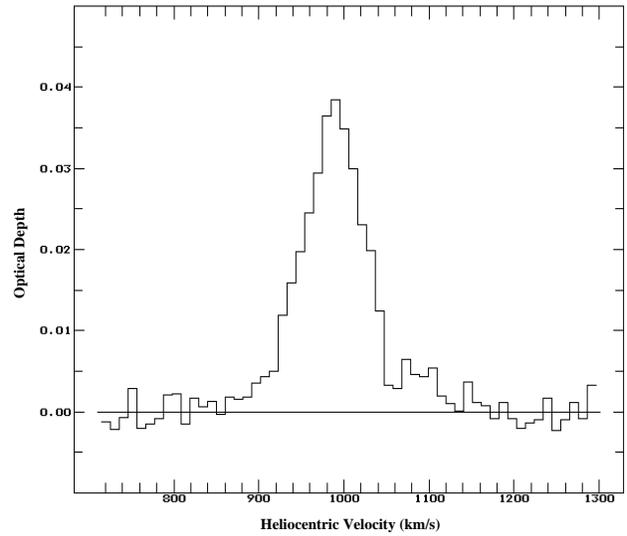}}
\end{picture} 
\caption{The average optical depth spectrum over the central beam area.}
\label{weightspec}
\end{figure}

The CLEANed continuum image was added back to the CLEANed
continuum-free cube, to provide the correct continuum level adjacent
to the line in each pixel. The absorption spectra coincident with the
nine brightest continuum pixels were selected to form the average
absorption spectrum.  A linear baseline was fitted and subtracted from
each spectrum before weighting each spectrum with its corresponding
continuum temperature. These weighted spectra were then averaged
together to produce the continuum-weighted average absorption
spectrum; the corresponding optical depth spectrum is shown in Figure
\ref{weightspec}. Table \ref{gaustab} shows the results of fitting two
Gaussian profiles to the absorption spectrum. It was possible to
obtain convergence, when fitting two Gaussians to the line, in two
different ways; (a) a deep narrow Gaussian centred at (987 $\pm$ 1) km
s$^{-1}$ and a smaller Gaussian redshifted to (1096 $\pm$ 6) km
s$^{-1}$; (b) a deep narrow component centred at 988 km s$^{-1}$ and a
broad shallow component centred at 986 km s$^{-1}$.  Pedlar et al.,
(1992) also note the presence of two components in their absorption
profile, centred at 988 and 1120 km s$^{-1}$. In both cases, when the
Gaussian components were subtracted from the present data, an emission
feature, with a peak of $\sim$1.4 mJy beam$^{-1}$, a FWHM of $\sim$20
km s$^{-1}$ and centred at $\sim$1050 km s$^{-1}$, was visible in the
residuals.

The absorption appears to be spatially unresolved at the resolution of
4.71$'' \times$ 4.71$''$, as the spectra do not show absorption
against the eastern extended continuum emission. This is consistent
with the absorption taking place on scales smaller than the B array
beamsize. In fact, Mundell et al. (1995b) studied this strong,
unresolved absorption at very high resolution (0.15$''$) with MERLIN
and even on these subarcsecond scales, the absorption, which is seen
only against the component which is thought to contain the AGN, is
marginally resolved.

\begin{table}
\centering 
\caption[Results of Gaussian fitting to the absorption profile]
 {Gaussian fits to the average absorption spectrum. Two Gaussian
 profiles were used each time to model the profile and two types of
 fit, (a) and (b) were equally applicable.}
\label{gaustab} 
\begin{tabular}{lcc} 
\hline & & \\ 
{\bf Line Centre} & {\bf FWHM} & {\bf Peak} \\ (km s$^{-1}$)& (km
      s$^{-1}$) &(mJy beam$^{-1}$)\\ (a) 987 $\pm$ 1 & 87 $\pm$ 3~ &
      6.6 $\pm$ 0.2 \\ (a) 1096 $\pm$ 6~ & 35 $\pm$ 15 & 0.9 $\pm$ 0.3
      \\
& & \\
(b)  988 $\pm$ 12	& 73 $\pm$ 23		& 5.6 $\pm$ 0.6   \\
(b) 986 $\pm$ 2	& 198 $\pm$ 6 ~		& 1.2 $\pm$ 0.6   \\
\end{tabular}
\end{table}

\subsubsection{High Velocity Cloud?}

\begin{figure}
\setlength{\unitlength}{1mm}
\begin{picture}(10,79)
 
\put(0,0){\includegraphics{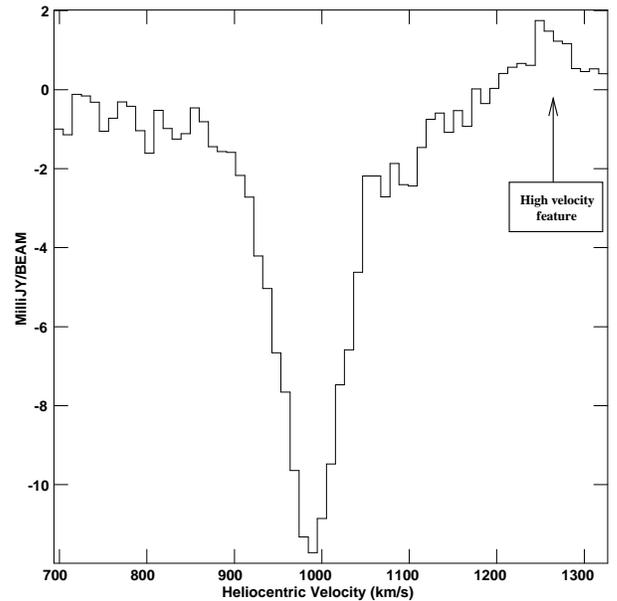}} 
\end{picture} 
\caption[Spectrum of high velocity feature] {Spectrum taken over the
central pixel of the nuclear region, showing the high velocity emission
feature to the left of the central absorption, centred at $\sim$1260
km s$^{-1}$.}
\label{speckntr}
\end{figure}

After subtraction of continuum emission, each channel map was examined
for signs of residual HI emission. The maps of these channels show a
compact region of emission, coincident with the nucleus, covering a
velocity range of 1249 km s$^{-1}$ to 1280 km s$^{-1}$. A spectrum
taken over the nuclear region (Figure \ref{speckntr}) shows the deep
absorption clearly and the emission feature close to the end of the
band, centred at $\sim$1260 km s$^{-1}$.  This feature may be
interpreted as a high velocity nuclear cloud, red-shifted by $\sim$260
km s$^{-1}$ with respect to systemic, with a column density of 9.6
$\times$ 10$^{20}$ cm$^{-2}$ and hence HI mass of 2.3 $\times$ 10$^7$
M$_\odot$. Since the cloud is coincident with the nucleus but does not
show up in absorption and has red-shifted velocities, it would
correspond to outflow on the far side of the nucleus.

As this signal is close to the edge of the band and is also spatially
coincident with the nucleus, it could be an artifact due to the strong
continuum emission from the nucleus. Broader band observations (or
indeed observations with the feature placed at the centre of the band)
are required to determine whether this feature is a real HI cloud or
if it is a baseline problem caused by the strong nuclear continuum
source.

\subsection{The HI Emission}

\begin{figure*}
\setlength{\unitlength}{1mm}
\begin{picture}(10,180)
 
\put(0,0){\includegraphics{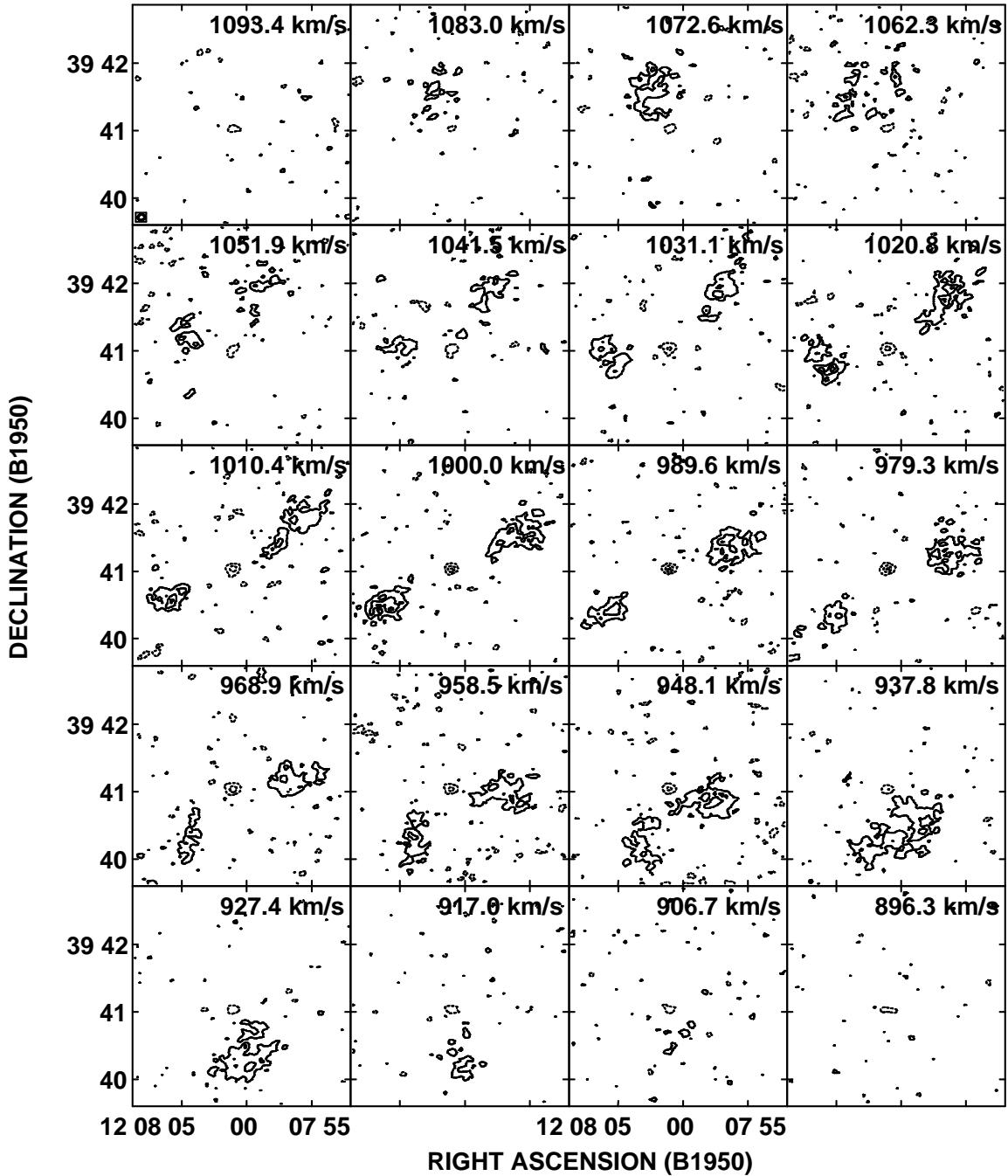}} 
\end{picture} 
\caption[B array channel maps of bar region in NGC4151]{Channel maps of HI emission from the central oval region, from the naturally weighted cube,  as in the previous
figure. The beam size is 6.07$'' \times$ 5.11$''$ and contour levels
are (-10, -5, -0.9, 0.9, 1.8, 2.7) mJy beam$^{-1}$, where 1 mJy
beam$^{-1}$ corresponds to N$_H$ = 3.73 $\times$ 10$^{20}$
cm$^{-2}$. Each panel is marked with its central velocity (km s$^{-1}$).}
\label{barchan}
\end{figure*}

Although the resolution of the uniformly weighted cube is higher than
that of the naturally weighted cube, making it useful for the
continuum and absorption study, the sensitivity to HI emission is
insufficient to perform a detailed analysis of the HI field; the
lowest detectable column density in the uniform cube is 2.8 x
10$^{20}$ cm$^{-2}$ compared with 1.32 x 10$^{20}$ cm$^{-2}$ in the
natural cube.  Therefore all subsequent analysis is performed on the
naturally weighted cube only.

The channel maps of those channels containing HI emission compare well
with those of Pedlar et al. (1992), with emission from both the bar
and outer spiral structure visible.  The spectra were well
approximated by single Gaussian profiles, which enabled moment
analysis to be performed on the data.  Maps of the channels containing
HI emission from the galactic bar are shown in Figure
\ref{barchan}. The emission covers the velocity range 907 km s$^{-1}$
to 1083 km s$^{-1}$. The absorption feature is also visible at the
centre of each map.

\subsection{HI Moment Maps}

\subsubsection{Method of Derivation}

Line integral profiles, or moments, can be calculated from a spectral
line cube. The zeroth moment corresponds to the integrated intensity
over velocity, the first moment to the intensity weighted velocity,
and the second moment to the intensity weighted velocity
dispersion. Several methods exist to derive moment maps of
galaxies. The methods range (in effort) from interactively fitting
Gaussian profiles to each spectrum at every pixel position across the
galaxy, to using a mainly automated routine such as MOMNT in AIPS.
Although the former method gives good results it is extremely time
consuming, especially for high resolution data, as a typical galaxy
may cover a grid of thousands of pixels (e.g. NGC4151 covers $\sim$300
$\times$ 300 pixels). The latter method is adequate for relatively
high signal-to-noise ratio data and to examine, quickly, the main
emission features in a galaxy.

\begin{figure*}
\setlength{\unitlength}{1mm}
\begin{picture}(50,220)
 
\put(0,0){\includegraphics{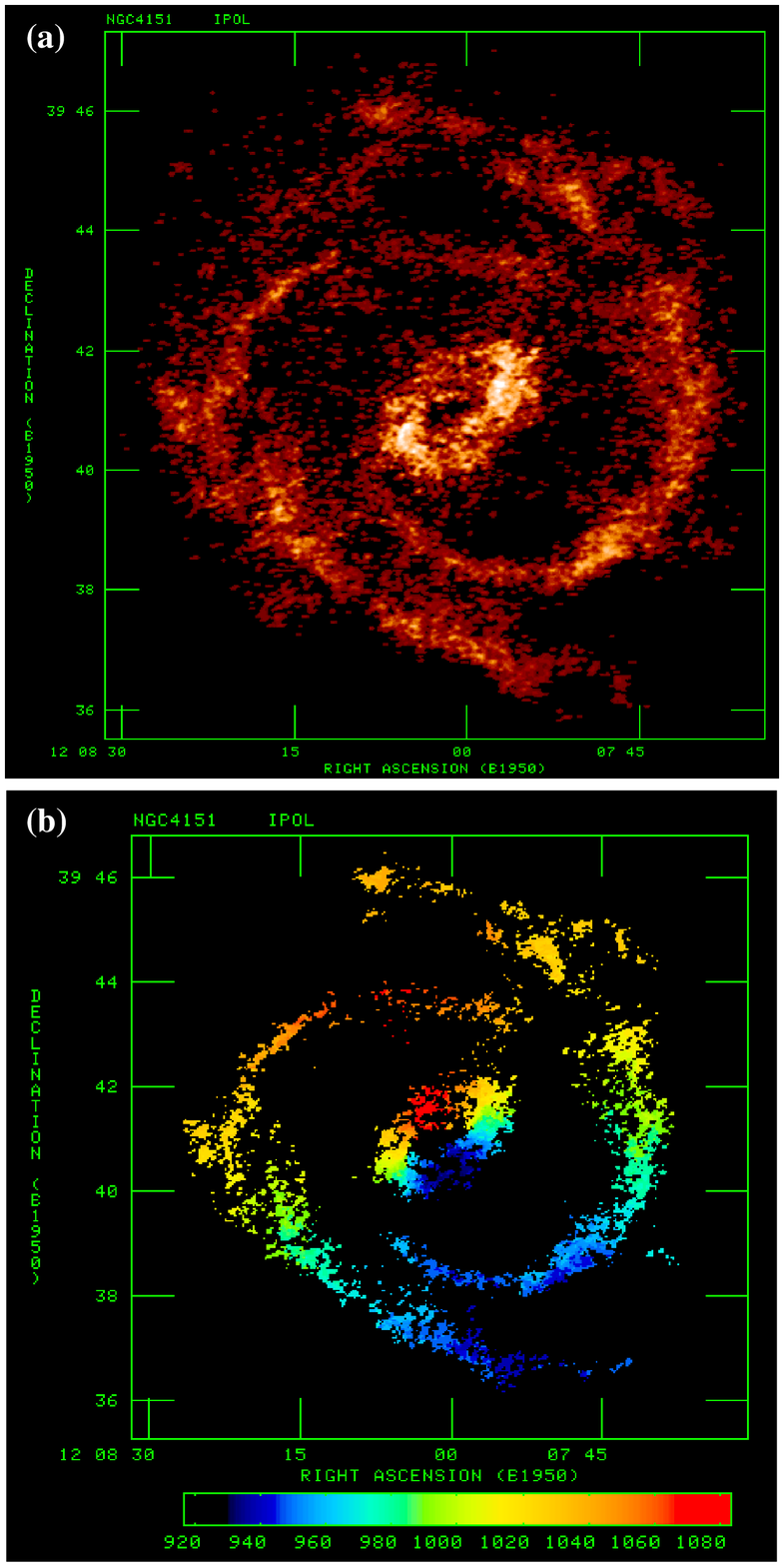}}
\end{picture}

\caption{(a) Zeroth moment map derived using the interactive moment
method. The flux cut-off level was 0.7 mJy beam$^{-1}$ and the
beamsize is 6.07$'' \times$ 5.11$''$; (b)
First moment map derived using the interactive moment method. The flux
cut-off level was 0.9 mJy beam$^{-1}$ and the resolution is the same
as in (a). The colour-velocity key is shown at the bottom with the
colours labelled in km s$^{-1}$.}
\label{momnt01}
\end{figure*}

The following method (Brinks, private communication) allows more of
the signal to be included in the moment analysis without including
more noise that will corrupt the analysis.  The method combines
features of the `cut-off' and `window' methods (Bosma, 1981). The
`cut-off' method involves summing (or analysing) all signals above a
specified noise limit, in each pixel, through the cube. Using this
method alone however can lead to the inclusion of noise spikes in the
final map. The `window' method involves examining each channel map and
interactively `marking' or drawing around `believable' regions of
emission and storing them for later use.  The window method has the
advantage over the cut-off method, for low signal to noise data, that
uncorrelated noise peaks, far away from the real signal and
uncorrelated in velocity, are more easily detected and discriminated
against (Teuben et al., 1986).
 
This combination of the two methods is clearly preferable to using
MOMNT, which is based on the `cut-off' method only so requires a
higher cut-off level to ensure that noise does not corrupt the
analysis. 
 
The B-array HI data cube was clipped so that only emission above
2$\sigma$ was included. Note that the absorption feature at the centre
is excluded by this method and hence  analysis of the absorption
structure was performed separately (see Section 4).  The clipped
cube was then examined interactively and `real' emission was
`windowed' off. Emission was defined to be `real' if it appeared in
more than one consecutive channel (i.e. continuity in velocity - at
least to a neighbouring channel) and was also present in the C array
data of Pedlar et al. (1992).

All emission outside of these windowed regions was blanked out as
noise. This blanked cube was then used as the `master' cube or
`template'. The original cube was  masked using the master cube
to produce the final blanked cube. XMOM was then used to perform the
moment analysis on this cube to produce the zeroth, first and second
moments from pixels that had survived the blanking process.

The resulting moment maps are shown in Figure \ref{momnt01}.  The B
array observations reveal structures that are remarkably similar to
those found in the previous C array observations, highlighting the
power of this type of moment analysis `by hand' for determining the
integrated neutral hydrogen distribution for apparently low
sensitivity observations. This method was found to be robust when
compared with conventional moment analysis (Thean, 1995).

\subsubsection{The Moment Maps}

The integrated neutral hydrogen image compares well with the features
in the optical image of Arp (1977) and the HI image of Pedlar et
al. (1992).

We derive an HI mass for the galaxy of 2.3 $\times$ 10$^9$ M$_\odot$,
of which 5.8~$\times$~10$^8$~M$_\odot$ is in the bar. These lower
limits compare well with those of Pedlar et al., (1992) of
3~$\times$~10$^9$~M$_\odot$ and 5~$\times$~10$^8$~M$_\odot$ for the
galaxy and bar respectively. Our estimate for the total HI mass is
slightly lower, probably due to some missing extended flux resulting
from a lack of short spacings, and the bar estimate is higher because
the improved resolution means that less of the emission in the bar was
lost to the absorption feature in the centre.

There are local minima of the HI column density in the outer regions
towards the NE and the SW, between the spiral arms and the bar, and
also inside the bar, approximately NW and SE of the central absorption
feature. These are real minima and not absorption features, as no
nuclear continuum emission is detected in these regions. They were
also noted by Pedlar et al., (1992) so are not due to our sensitivity
limitations. The density minima along the bar minor axis, between the
bar and spiral arms, may be explained by the presence of unstable
periodic orbits around the Lagrangian points, L$_4$ and L$_5$
(i.e. points at co-rotation, along the bar minor axis) (Contopoulos,
1981; Schwarz, 1984); the lack of stable periodic orbits means that
non-periodic orbits have no centres of attraction, so cannot be
trapped, thereby causing large stochasticity and depopulation of these
regions (Schwarz, 1981; Contopoulos, 1981).

The first moment (Figure \ref{momnt01}(b)), representing the intensity
weighted velocity field, is consistent with that of Pedlar et
al. (1992) and is discussed further in Section 3.6.

The second moment corresponds to the intensity weighted dispersion so
gives an indication of the variation in line-widths across the
galaxy\footnote{For a Gaussian line profile the Full Width at Half
Maximum is 2$\sqrt{2ln2}$ $\sigma$ where $\sigma$ is the second
moment.}. The dispersion values in NGC4151 range from $\sim$5 to 25
km s$^{-1}$; most of the galaxy has dispersions of $\sim$5 km
s$^{-1}$, typical of galactic values (dispersions of 5-8 km s$^{-1}$--
Dickey, Hanson \& Helou, 1990), while slightly higher dispersions (10 km
s$^{-1}$) are seen in the outer spiral arms in the regions of enhanced
emission. The largest dispersions (up to $\sim$25 km s$^{-1}$) occur
in the bar shocks (see Paper I).

\subsection{The Outer Spiral Structure}

The two outer spiral arms of NGC4151 are clearly visible in the
present observations and can be traced out to $\sim$5.5$'$ (21.5 kpc)
from the centre, as in the Pedlar et al., (1992) image. The structure
shows all of the features visible in the lower resolution data (Pedlar
et al., 1992); e.g. the region of enhanced emission that lies 3.5$'$
SW of the nucleus in the eastern spiral arm (which is also a location
of enhanced optical emission).  

\subsubsection{An HI Signature of Seyferts?}

As noted by Pedlar et al., the spiral arms in NGC4151 are particularly
gas-rich but show little evidence of star formation.  Emission from
the central oval dominates optical images of the galaxy, but in the HI
the column densities in the spiral arms are comparable to those in the
bar. Although the optical emission in the spiral arms is faint, deep
optical images (Arp, 1977) show that the optical emission and HI
distribution are closely matched. There appears to be no HI emission
beyond the optical extent of the galaxy. This is in contrast to many
normal galaxies in which, more usually, the HI extends beyond the
optical (Bosma, 1983, Sancisi, 1983, Wevers, 1984, Broeils \& Rhee,
1997). Other Seyferts show a similar effect (e.g., NGC1068, Brinks et
al. 1994, Brinks \& Mundell 1997; NGC4051, Liszt \& Dickey; NGC4939,
Mundell 1997; NGC5033, Thean et al. 1997), in which the HI and optical
emissions are coincident. The only Seyferts in which the HI extends
beyond the optical structure (e.g., Mkn348, Simkin et al., 1987;
NGC3227, Mundell et al. 1995a; NGC3982, Mundell 1997; NGC5506, Mundell
1997) show clear evidence of tidal disturbance. Alternatively, the
close correspondence between the HI and optical extent may be more
typical of early-type barred galaxies; studies of NGC1433 (Ryder et
al. 1996) and NGC5850 (Higdon et al., 1998) show a close match between
the HI and optical structures.

Detailed HI synthesis studies and deep optical imaging of a larger
sample of Seyferts and normal galaxies would be required to determine
whether this is an HI property specific to Seyferts.

\subsubsection{Evidence for Interaction?}

Pedlar et al., (1992) note that the spiral structure appears
relatively undisturbed and rule out the possibility that NGC4151 has
been involved in a (strong) interaction with any of its
neighbours. However, the present observations show some evidence of
distortion from a symmetrical two-armed spiral pattern and possible
evidence of tidal disturbance.

\begin{figure}
\setlength{\unitlength}{1mm}
\begin{picture}(50,65)
 
\put(0,0){\includegraphics{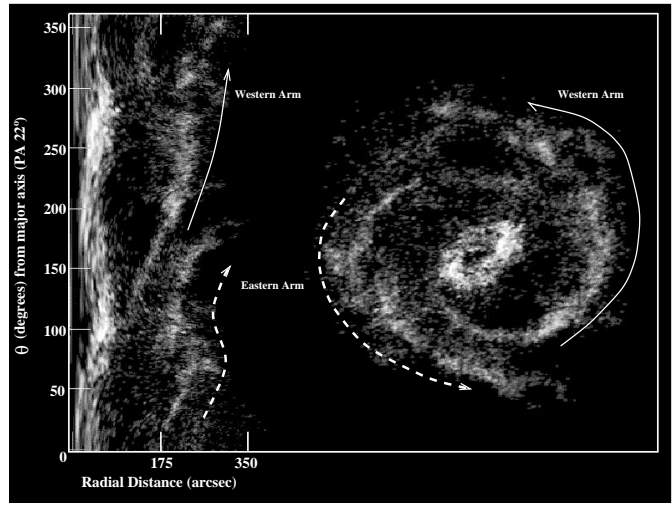}}
\end{picture} 
\caption[The spiral structure of NGC4151 in polar co-ordinates]{The spiral structure of NGC4151, on the right, is shown in polar co-ordinates on the left. The vertical axis represents azimuthal angle ($\theta^{\circ}$),
measured anticlockwise from the major axis in PA 22$^{\circ}$. The
linear scale on the horizontal axis represents distance from the centre
of the bar. Each segment of the spiral arms can be easily traced in
the polar diagram, as indicated by the dotted lines.}
\label{polar}
\end{figure}

Firstly, the arm segments that lie in the upper half of the image seem
to have been displaced upwards. The eastern arm (see Figure
\ref{polar}) is disconnected from the end of the bar compared with the
western arm.  A more quantitative examination of the spiral pattern
can be performed by using the polar diagram in Figure \ref{polar}. In
this image the emission is plotted as a function of azimuthal angle,
$\theta$ (measured east from the major axis in PA 22$^0$), against
distance from the centre of the galaxy. A circle would be represented
by a vertical line in the polar plot. The pitch angle of the spiral at
any radius r, defined as the tangent to the spiral arm and the tangent
to the circle (r = constant) (Binney \& Tremaine, 1987), can be easily
measured from the polar diagram.  Therefore the angle made by the
spiral feature and the vertical axis is the pitch angle.

\begin{figure}
\setlength{\unitlength}{1mm}
\begin{picture}(50,135)
 
\put(0,0){\includegraphics{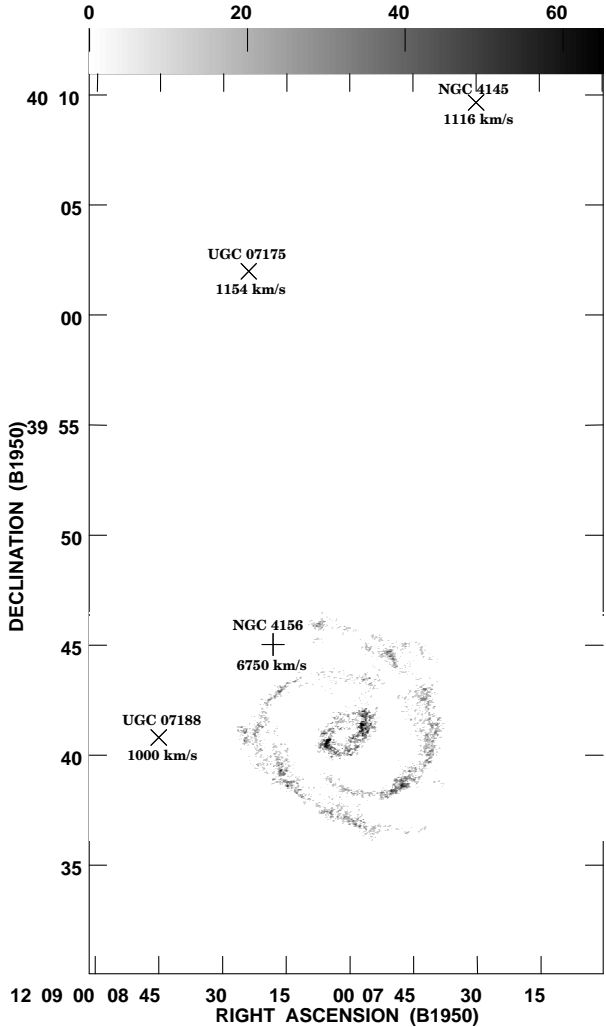}}
\end{picture} 
\caption[Nearby neighbours of NGC4151]{The neutral hydrogen distribution of NGC4151 and the positions of three nearby galaxies (marked by crosses), noted by Bosma et al., (1977), with systemic velocities close to that of NGC4151. NGC4156 is also marked, despite its higher systemic velocity. The systemic velocity of each galaxy is also indicated} 
\label{inters}
\end{figure}

 The pitch angle of the western arm remains constant at
$\sim$22$^{\circ}$ in the inner regions (i.e. from 100$^{\circ}$ $<$
$\theta$ $<$ 220$^{\circ}$ or PA of 122$^{\circ}$ to 242$^{\circ}$),
close to the southern end of the bar but then begins to decrease, (at
$\theta$ $\sim$220$^{\circ}$), to $\sim$3$^{\circ}$.  Each arm also is
`broken' at about half way along its bend. The western arm is broken
but the broken segment appears to follow round continuously, albeit it
at a reduced pitch angle. The eastern arm, however, is broken at
$\theta\sim$80$^{\circ}$ (PA 102$^{\circ}$), with the two segments
appearing shifted or `dislocated' from one another. The pitch angle
changes from $\sim$27$^{\circ}$ for the segment that starts from the
northern end of the bar, to $\sim$16$^{\circ}$ for the second segment
that leads to the outer tip of the arm.

Figure \ref{inters} shows NGC4151 and some neighbouring galaxies
(Bosma et al., 1977) which have systemic velocity values close to that
of NGC4151. NGC4156, which is also marked, has been suggested by Arp
(1977) to be interacting with NGC4151 despite its rather larger value
of systemic velocity of 6750 km s$^{-1}$. NGC4145, despite its more
suitable velocity, seems to be rather far from NGC4151 to be
responsible for the observed disturbance. UGC07175 seems another
possible candidate along with UCG07188 but interaction simulations
are needed to determine whether any of these galaxies could have been
responsible for tidally disturbing NGC4151.

\subsection{The HI Rotation Curve of NGC4151}

To derive the rotation curve of a spiral galaxy to first order, one
may fit a curve such as an exponential (linearly increasing to start
with then flat thereafter) or Brandt curve (attains a maximum value
then declines as Keplerian thereafter) to the velocity field of the
galaxy. However, real rotation curves deviate from mathematical models
due to warps, deviations from circular orbits and irregular mass
distributions. Therefore after using a mathematical model of the whole
galaxy velocity field to determine the best starting input parameters,
it is better to determine the galaxy parameters by fitting curves of
constant velocity to a series of concentric annuli. Generally, the
radial velocity, V$_r$, at some azimuthal angle, $\theta$, on an
annulus of radius r can be expressed as V$_r$ = V$_s$ + V$_c$ sin i
cos $\theta$, where V$_c$ is the circular velocity of the annulus and
V$_s$ is the systemic velocity of the galaxy.  So, for an annulus of
radius r, the velocity will vary sinusoidally with angle $\theta$
around the annulus, measured from the line of nodes. V$_c$ and i are
coupled (Begeman, 1985) but by keeping i fixed, one can solve for
V$_c$ and V$_s$, assuming that V does not change across the annulus.

\begin{figure}
\setlength{\unitlength}{1mm}
\begin{picture}(50,120)
 
\put(0,0){\includegraphics{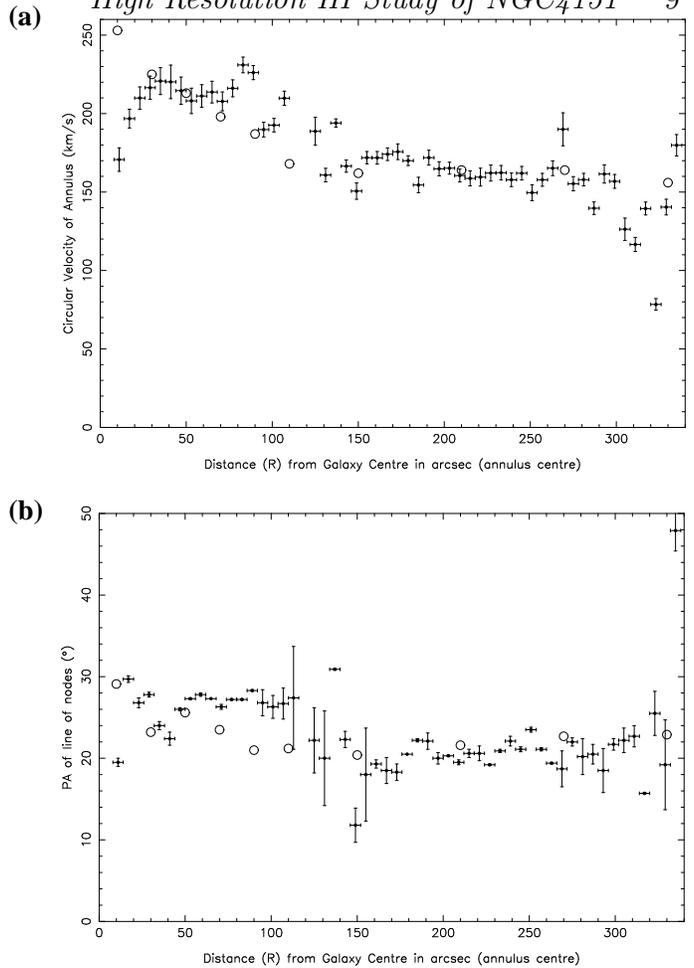}}
\end{picture} 
\caption{(a) The HI rotation curve of NGC4151 derived by fitting constant
velocity curves to a series of concentric annuli of thickness
6$''$. The open circles represent the rotation curve as derived by
Pedlar et al. (1992) (b) The change in the position angle of the line
of nodes as a function of distance from the centre of the galaxy. Note
the clear offset in the line of nodes associated with the bar. Again,
the open circles represent the data of Pedlar et al.}

\label{rotcur}
\end{figure}

The rotation curve of NGC4151 was derived from the first moment map
using this `tilted' ring model. The width of the annulus (6$''$) was
optimised such that it was sufficiently narrow to assume constant
velocity across it and wide enough that it contained enough points to
enable convergence of the fit.  Assuming that all rings were coplanar
(i.e., neglecting any warping), the inclination was held fixed at
21$^{\circ}$ throughout the fitting, but the PA of the line of nodes,
the co-ordinate of the annulus centre and the systemic velocity were
allowed to vary.  Figure \ref{rotcur}(a) shows the resulting rotation
curve derived from the B array observations. The first annulus started
at a radius of 8$''$ from the centre to avoid the central absorption
feature and high velocity cloud.  Also shown in Figure \ref{rotcur}(a)
is the rotation curve derived by Pedlar et al. (1992) from their lower
resolution data. The two compare well but the  rotation curve
from the new high resolution data shows a turnover at a radius of
$\sim$35$''$ which was not evident  in the data of Pedlar
et al., due to the lower resolution of their observations and the
contamination from the central absorption feature.

To first order, the assumption of circular orbits is valid but, as can
be seen in Figure \ref{rotcur}(b), there are clearly non-circular
motions associated with the bar potential as the PA of the line of
nodes changes with distance from the centre.  The rotation curve will
be used later to investigate the possible existence and positions of
resonances in the galaxy (See Section 4).

\section{Discussion}

A primary aim of detailed studies of barred galaxies is to ascertain
such physical properties as the global mass distribution, angular
velocity (pattern speed) and degree of central mass concentration of
the bar; properties which are not directly observable. However, the
general theoretical understanding relating such physical properties of
bars to observable phenomena (such as shocks, dust lanes, rings) has
become well established in recent years (e.g. Athanassoula, 1992a,b;
Lindblad, Lindblad \& Athanassoula, 1996).

Analytical and numerical works have predicted a variety of observable
phenomena, which, when detected, allow useful constraints to be placed
on the physical properties of the bar (e.g. Sanders \& Tubbs, 1980;
Van Albada \& Roberts, 1981; Combes \& Gerin, 1985; Athanassoula,
1992a,b; Sellwood \& Wilkinson, 1993, and references therein).  The
gas dynamics of the bar in NGC4151 are discussed in detail in Paper I
(Mundell \& Shone, 1998); here we relate the bar dynamics to the global
properties of the galaxy.  In particular, the characteristics of the
streaming shocks in NGC4151 provide valuable information about the
types and properties of the different families of orbits present in
the bar, which in turn are determined by the existence and location of
resonances in the galaxy (Teuben et al., 1986).

\subsection{Bar pattern speed and ILRs}

\begin{figure}
\setlength{\unitlength}{1mm}
\begin{picture}(45,60)
 
\put(0,0){\includegraphics{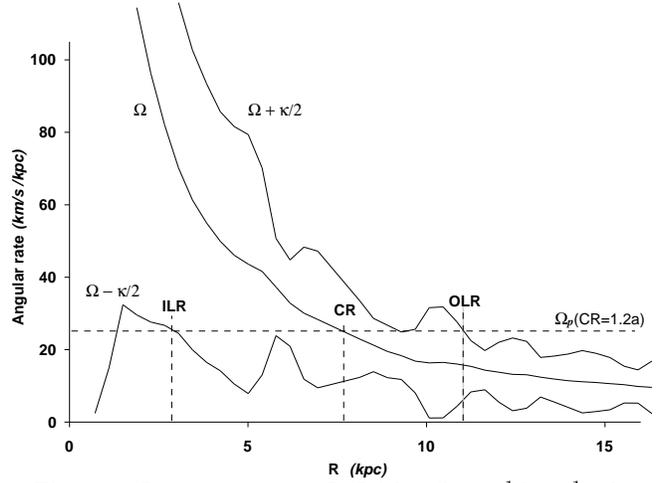}}
\end{picture} 
\caption[Resonance curves of NGC4151]{Resonance curves, $\Omega$,
$\Omega \pm k/2$ (km s$^{-1}$ kpc$^{-1}$) plotted against radial
distance from the centre of the galaxy, R (kpc).}
\label{omega}
\end{figure}

The new HI rotation curve of NGC4151 (Figure \ref{rotcur}(a) was
smoothed (using a box car average of 10 km s$^{-1}$) and used to
produce the resonance curves from derived values of
$\Omega$~=V~(km~s$^{-1}$)/R (kpc) and
\[k=(2{\Omega}){\sqrt{{(1+\frac{R}{2{\Omega}}} \frac{d{\Omega}}{dR}}).}\]
The derived quantities, $\Omega$, $\Omega \pm k/2$, are plotted as a
function of radial distance from the centre of the galaxy (defined by
the position of the radio continuum peak) in Figure \ref{omega}.

In order to determine the bar pattern speed and ILR locations, the
position of the CR is required.  As described in Paper I, simulations
of gas flows in weak bars (Athanassoula, 1992b) show that in order to
produce such offset shocks as those present in NGC4151, co-rotation is
situated just beyond the end of the bar, at a radius of
(1.2$\pm$0.1){\em a} (where {\em a} is the semi-major axis of the
bar). Table \ref{tabbar} shows the corresponding pattern speed of the
bar and location of the ILRs for three different positions of CR,
derived from the curves in Figure \ref{omega}. Although the
$\Omega-k/2$ shows a small turnover at the inner most radii, the
uncertainty in the amplitude here is such that more data points closer
to the nucleus are required to confirm whether this is truly a
turnover, which will then result in a second ILR.

It should be noted that although the above resonance treatment is a
valid first order approximation, the presence of significant
non-circular motions and non-linear shock regions in the bar means
that it is not possible to derive the radial mass distribution from
the rotation curve alone. Full simulations of the gas response to
potentials, constrained by additional observed properties presented
here, are required (Shone \& Mundell in preparation). 

\begin{table}
\centering 
\caption{Possible bar pattern speeds ($\Omega_p$) and Inner Lindblad
Resonance  radii for different chosen values of Co-Rotation radius
(CR), where {\em a} (99$''$) is the bar semi-major axis.}
\label{tabbar} 
\begin{tabular}{lcc} 
\hline
{\bf CR } & {\bf $\Omega_p$} & {\bf ILR} \\ &(km s$^{-1}$ kpc$^{-1}$)&
	  (kpc)\\ 1.1 {\em a} & 28.2 & 2.2 \\ 1.2 {\em a} & 24.5 & 2.8 \\ 1.3
{\em a} & 20.8 &
	  3.4 \\ \hline
\end{tabular}
\end{table}

\subsection{Bar-Driven Spiral Pattern in NGC4151?}

A simple model (Sanders \& Huntley, 1976) in which a potential due to
a rigidly rotating bar is introduced to a uniform disk of gas, which
has no self-gravity and initially circular orbits, demonstrated that a
strong, trailing spiral pattern is produced when the gas reaches
steady state. The gas response, to even a weak bar, is spiral -- even
though the forcing potential is not -- and strong, with a
density enhancement of as much as a factor of 2.5 in the spiral arms
(Binney \& Tremaine, 1987).  The spiral forms as a consequence of
viscosity in the gas and an open spiral structure, such as that in
NGC4151, requires only a bar and a dissipative ISM.  This bar-driven
spiral model would seem promising for NGC4151 with its weak bar
potential and greatly enhanced HI spiral arms.

The spiral arms in NGC4151 (as suggested by the de Vaucouleurs
classification) are similar to pseudo-rings commonly observed in SB
and SAB galaxies (Buta \& Crocker, 1991) and are classified as type
R$_2'$. This classification describes the $\sim$270$^{\circ}$ winding
of the spiral arms with respect to the ends of the bar, which results
in a doubling of the spiral arms in the two opposing quadrants
trailing the bar. The arms clear the opposite side of the bar
completely, thus distinguishing it from the more ring-like R$_1'$
morphology.  Such R$_2'$ pseudo-rings may be the response of galactic
disk gas to forcing by a bar (Schwarz, 1981). In this case the OLR is
found to lie close to the minor axis of the pseudo-ring, which for
NGC4151 would be at a radius of $\sim$16.5 kpc. Constraining the OLR
radius to this value and assuming that the spiral and bar pattern
speeds are equal, would result in a CR radius of $\sim$9 kpc and a
pattern speed of $\sim$18 km s$^{-1}$.

This CR radius is larger than the observed bar extent in NGC4151, but
Schwarz (1981) notes that although bars end at CR, the maximum of the
bar may occur at the 4:1 resonance, some way inside CR. However, the
bar-driven spiral models also predict that the spiral pattern extends
from CR to beyond the OLR; in NGC4151 the spiral pattern, although
asymmetric, begins just beyond the ends of the HI bar, which suggests
that CR is located here (as assumed in Section 4.1). Therefore, using
the CR radius and pattern speed derived in Section 4.1 and assuming
the bar and spiral patterns speeds to be the same, the OLR would be
situated at a radius of 10-11 kpc, implying that the spiral pattern
extends well beyond the OLR.

Alternatively, Sellwood \& Wilkinson (1993) point out that there is no
direct observational evidence that spiral patterns are driven by their
bars, and instead, it is often suggested that the spiral arms have lower
pattern speeds than the bar.  This may indeed be the case in NGC4151
where, if we assume the spiral pattern to finish at the Outer Lindblad
Resonance and CR to lie just beyond the end of the HI bar, we derive a
pattern speed for the spiral of $<$15~km~s$^{-1}$ compared with $\sim$25
km s$^{-1}$ for the bar. 

\subsection{Closer to the nucleus}

Theoretical and numerical simulations have shown that galactic bars
can drive large inflows of gas towards the nucleus, creating or
enhancing a central mass concentration, and possibly holding important
implications for the fuelling of AGN and galactic evolution generally
(e.g. Simkin, Su \& Schwarz, 1980, Shlosman, Frank \& Begelman, 1989,
Teuben 1996 and references therein). In the lower resolution HI study
of Pedlar et al. (1992), the HI in NGC4151 appeared as a continuous
neutral hydrogen feature crossing the nucleus, which they suggested
was a bridge of HI, originating in the shocks, continuing across the
nucleus and resulting in the HI absorption against the radio continuum
nucleus. However, further investigation of both the HI emission and
absorption at higher angular resolution reveals this interpretation to
be unlikely.

The high resolution (0.16$''$ or 10 pc) MERLIN absorption study
(Mundell et al., 1995b) revealed that the HI absorption did not take
place against the entire length of the 4$''$-long radio continuum jet,
but instead was localised to the component (C4) which is thought to
contain the active nucleus. The absorption was attributed to a 90
M$_\odot$ cloud of HI, 1.5pc in extent, which may form part of a
nuclear obscuring torus as advocated in Unification Schemes (Antonucci
\& Miller, 1985; Mundell et al., 1995b).

The high angular resolution HI emission data presented in this paper
allows us to resolve individual features in the bar for the first time
(e.g. Mundell \& Shone, 1998). These data show that the HI does not
stream directly from the shocks onto the nucleus but instead loses
angular momentum in the shocks and moves onto smaller orbits closer to
the nucleus, forming fingers of HI which wind {\em around} the
nucleus. In addition, the HI fingers correspond closely with two
arc-like features discovered in the optical colour index map of
Vila-Vilaro, (1995). These two red arcs lie along opposite
quadrants of an ellipse, which has a semi-major axis of 18$''$ and is
approximately perpendicular to the bar major axis. Vila-Vilaro et
al. attribute the ring to dust extinction of the normal stellar
continuum of the galactic bulge, an interpretation supported by the
discovery of polarisation due to dichroic absorption around the
ellipse by Draper et al. (1992).

Studies of orbits in non-axisymmetric potentials and the resulting gas
flows, which predict the formation of shocks, rings and nuclear
spiral-like features similar to those seen in NGC4151, help to clarify
why the simplistic interpretation of a continuous bridge of HI
crossing the nucleus of NGC4151 is not applicable. The shocks in the
bar provide direct evidence of large-extent $x_2$ (perpendicular)
orbits in the bar, without which, such curved offset shocks could not
form (Athanassoula, 1992b). The dust ellipse may be associated with gas
settling onto the $x_2$ orbits, as in the case of M100 (Knapen et al.,
1995). The condition (Athanassoula, 1992b) that at least one ILR is
required for the $x_2$ family to exist (and if two ILRs exist, the
$x_2$ orbits must lie between them), is satisfied in the case
of NGC4151, with the dust ellipse lying inside the ILR, whichever of
the three positions of CR is used. If a second ILR does exist it must
lie closer to the nucleus and therefore would not be detected in the
present HI data.

\subsubsection{Comparison with M100}

Tightly wound spirals or ring-like features have been observed in M100
(Knapen et al., 1995), a grand design two-armed spiral galaxy with a
central bar and mild star-formation activity in the nuclear regions.
Self-consistent, three dimensional numerical models, with and without
star-formation, were used to understand the circum-nuclear structures
in M100 (Heller \& Shlosman 1994, Knapen et al. 1995). In the model
without star-formation, a large-scale pair of leading-edge, trailing
shocks formed, consistent with simulations by Athanassoula (1992) and
similar to NGC4151 (Mundell \& Shone, 1998). In addition, a pair of
tightly wound trailing shocks and a smaller pair of leading shocks
formed in the central 2.5 kpc of M100. These two shock systems are
shown to interact non-linearly, forming two gas circulations, or oval
rings, between the two ILRs in the bar (Shlosman, 1996). Evolution of
the system, with self-gravity of the gas included, then shows the
outer gaseous ring to settle down onto the the lower energy $x_2$
orbits, inside the OILR, and the inner ring to shrink across the IILR,
settling onto smaller $x_1$ inside the IILR (Shlosman, 1996). 

We speculate that a similar situation may be applicable to NGC4151,
with the Vila-Vilaro et al. (1995) dust ellipse corresponding to the
M100 outer ring, lying inside the OILR and associated with the $x_2$
orbits in that region.  Although an obvious counterpart to the
M100-inner ring is not seen in NGC4151, a number of interesting
features do exist on smaller scales such as a small blue ring,
approximately 2--3$''$ in extent (Kotilainen \& Ward, 1997) and a red
compact structure approximately 1 -- 2$''$ long, oriented
approximately along the main HI bar (Perez et al., 1989; Terlevich et
al., 1991). Suggestions that the red feature could be due to the
presence of an obscuring torus (Perez et al., 1989; Terlevich et al.,
1991), which collimates the UV radiation responsible for producing the
ENLR (Robinson et al., 1994), would imply that the properties of the
active nucleus may be directly related to gaseous conditions in the
bar of the host galaxy (Axon \& Robinson, 1996). However, high angular
resolution optical and infra-red kinematical studies, tracing the
rotation curve closer into the nucleus, and determining the kinematics
of the nuclear structures, combined with fully self-consistent
three-dimensional numerical simulations of the bar would be required
to determine the existence of a second ILR and the true nature of the
nuclear optical features; contamination from NLR emission however,
does cause significant problems in separating rotation from outflow
velocities.

\section{Conclusions}

\begin{itemize}

\item Sensitive, high angular resolution ($6''\times5''$ -- 390 pc)
$\lambda$21-cm observations of the neutral hydrogen in the nearby
barred Seyfert galaxy, NGC4151, have been presented. 

\item On the smallest scales, spatially unresolved HI absorption (at a
resolution of 4.71$''$) against the radio continuum nucleus is
detected. Gaussian fits to the two-component absorption spectrum
converge with either a deep narrow Gaussian centred at (987 $\pm$ 1)
km s$^{-1}$ and a smaller Gaussian redshifted to (1096 $\pm$ 6) km
s$^{-1}$, or a deep narrow component centred at 988 km s$^{-1}$ and a
broad shallow component centred at 986 km s$^{-1}$

\item A high velocity cloud is detected in emission, coincident with
the nucleus. If this feature is not due to baseline problems, it
indicates the presence of a cloud with an HI mass of 2.3 $\times$
10$^7$ M$_\odot$ and red-shifted by 260 km s$^{-1}$ from systemic,
corresponding outflow on the far side of the nucleus.

\item The HI bridge which was suggested, from lower resolution
studies, to originate in the shocks and cross the nucleus directly, is
not detected in the present study. Instead, the gas streaming from the
shocks forms fingers of HI which wind {\em around} the nucleus.

\item The HI fingers correspond closely with dust arcs seen in optical
studies; these features are consistent with predictions from general
numerical simulations of bars, and resemble nuclear features seen by
others in weak barred galaxies such as M100.

\item The new rotation curve presented extends to within 8$''$ of the
nucleus and shows a turnover at a radius of $\sim$35$''$, which was
previously undetected in lower resolution studies.  

\item The derived resonance curve and the properties of the shocks
(Paper I) yield a bar pattern speed of 24.5 km s$^{-1}$ $\pm$3.7 km
s$^{-1}$ and one (ILR) at a radius of 2.8 $\pm$0.6 kpc.  If a second
ILR exists closer to the nucleus it would not be detected in the
present HI data.

\end{itemize}

\section{Acknowledgements}

We thank Elias Brinks for useful discussion and the referee, Professor
Ron Buta for helpful suggestions. CGM acknowledges a research
fellowship from the U.K. Particle Physics and Astronomy Research
Council (PPARC).

{}

\end{document}